\documentclass[twoside,australian,sort&compress]{iopart}
\usepackage[T1]{fontenc}
\usepackage[utf8]{inputenc}
\usepackage{geometry}
\usepackage{graphicx}
\usepackage[numbers]{natbib}

\makeatletter
\usepackage{iopams}
\usepackage{setstack}

\newcommand{\eqref}[1]{(\ref{#1})}

\usepackage{xurl}

\makeatother

\usepackage{babel}
\begin{document}
\title{Determining cross sections from transport coefficients using deep
neural networks}
\author{{\Large{}P W Stokes$^{1}$, D G Cocks$^{2}$, M J Brunger$^{3,4}$
and R D White$^{1}$}}
\address{{\Large{}$^{1}$}{\large{}College of Science and Engineering, James
Cook University, Townsville, QLD 4811, Australia}}
\address{{\Large{}$^{2}$}{\large{}Research School of Physics, Australian National
University, Canberra, ACT 2601, Australia}}
\address{{\Large{}$^{3}$}{\large{}College of Science and Engineering, Flinders
University, Bedford Park, Adelaide, SA 5042, Australia}}
\address{{\Large{}$^{4}$}{\large{}Department of Actuarial Science and Applied
Statistics, Faculty of Business and Information Science, UCSI University,
Kuala Lumpur 56000, Malaysia}}
\ead{{\large{}peter.stokes@my.jcu.edu.au}}
\begin{abstract}
We present a neural network for the solution of the inverse swarm
problem of deriving cross sections from swarm transport data. To account
for the uncertainty inherent to this somewhat ill-posed inverse problem,
we train the neural network using cross sections from the LXCat project,
paired with associated transport coefficients found by the numerical
solution of Boltzmann's equation. The use of experimentally measured
and theoretically calculated cross sections for training encourages
the network to avoid unphysical solutions, such as those containing
spurious energy-dependent oscillations. We successfully apply this
machine learning approach to simulated swarm data for electron transport
in helium, separately determining its elastic momentum transfer and
ionisation cross sections to within an accuracy of $4\%$ over the
range of energies considered. Our attempt to extend our method to
argon was less successful, although the reason for that observation
is well-understood. Finally, we explore the feasibility of simultaneously
determining cross sections of helium using this approach. We have
some success here, determining elastic, total $n=2$ excitation and
ionisation cross sections to $10\%$, $20\%$ and $25\%$ accuracy,
respectively. We are unsuccessful in properly unfolding the separate
$n=2$ singlet and triplet excitation cross sections of helium, but
this is as expected given their similar threshold energies.
\end{abstract}
\noindent{\it Keywords\/}: {swarm analysis, inverse problem, Boltzmann equation, machine learning\\
}
\submitto{\PSST }
\maketitle

\section{\label{sec:Introduction}Introduction}

Control and optimisation of plasma processing is dependent on accurate
modelling of the associated charged particle transport. Fundamental
to this is the provision of complete and accurate sets of cross sections.
These cross sections must have embedded particle, momentum and energy
balance built into them. Experiment (e.g. crossed beam \citep{Filippelli1994})
and theory (e.g. convergent close-coupling \citep{Bray1992,Fursa1995,Zammit2014,Zammit2016})
can provide some of this required information however there is often
still uncertainty associated with these cross-sections (e.g. where
extrapolation is required, or where theories break down) and here
``educated guesses'' are often used. Swarm experiments provide accurate
yet independent data to assess the self-consistency of these cross
sections \citep{White2018}, which in principle makes them very useful
to evaluate their utility.

In this study, we assess the inference/extraction of microscopic information
from measured swarm data. The first attempts at deriving electron
scattering cross sections from swarm transport coefficients were made
in the 1920s by Mayer \citep{Mayer1921}, Ramsauer \citep{Ramsauer1921}
and Townsend \textit{et al.} \citep{Townsend1922}. Early approaches
such as these assumed a simplified form of the electron energy distribution
function (EEDF), such as a Maxwellian or Druyvesteyn distribution,
from which theoretical transport coefficients could be calculated
and contrasted with those measured experimentally. By iteratively
adjusting the cross sections until the calculated transport coefficients
coincided with those from experiment, a plausible solution to this
\textit{inverse swarm problem} could be found. The accuracy of this
iterative approach was improved in the 1960s when Phelps, alongside
numerous collaborators, began determining the EEDF from the numerical
solution of Boltzmann's equation \citep{Frost1962,Engelhardt1963,Engelhardt1964,Hake1967,Phelps1968}:
\begin{equation}
\left(\frac{\partial}{\partial t}+\mathbf{v}\cdot\frac{\partial}{\partial\mathbf{r}}-\frac{e\mathbf{E}}{m}\cdot\frac{\partial}{\partial\mathbf{v}}\right)f\left(t,\mathbf{r},\mathbf{v}\right)=\left(\frac{\partial f}{\partial t}\right)_{\mathrm{coll}},\label{eq:BE}
\end{equation}
where $f\left(t,\mathbf{r},\mathbf{v}\right)$ is the phase-space
distribution function of the electron swarm, $e$ is the elementary
charge, $m$ is the electron mass, $\mathbf{E}$ is the applied electric
field, and $\left(\frac{\partial f}{\partial t}\right)_{\mathrm{coll}}$
contains operators for all the considered electron collision processes.

\begin{figure}
\begin{centering}
\includegraphics[scale=0.6]{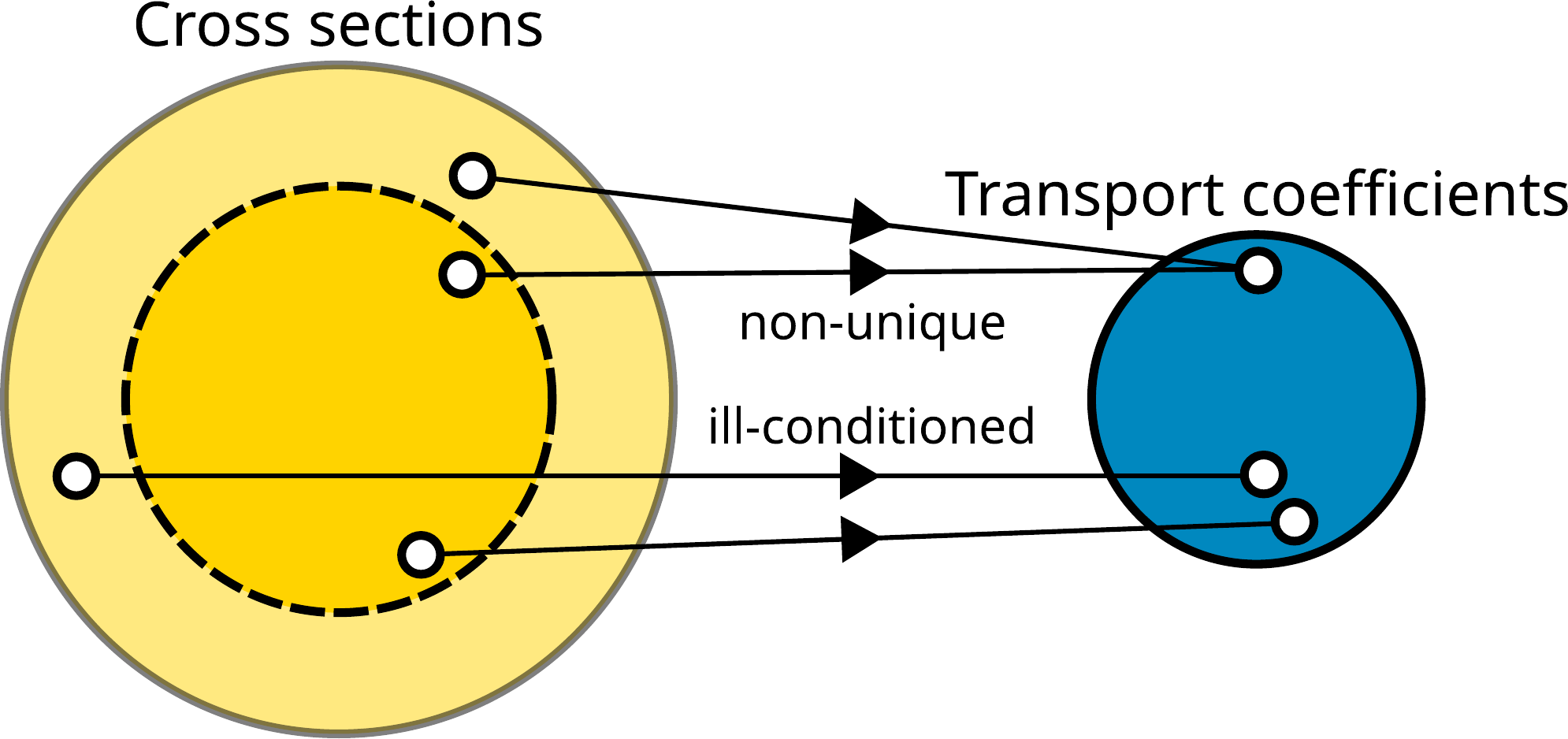}
\par\end{centering}
\caption{\label{fig:Euler-diagram-illustrating}Euler diagram illustrating
the somewhat ill-posed nature of the inverse swarm problem. Arrows
denote the forward problem of mapping from a set of cross sections
to a set of corresponding transport coefficients. This mapping is
well-posed and can be achieved numerically by solving Boltzmann's
equation \eqref{eq:BE} or by performing a Monte Carlo simulation,
among other techniques. Conversely, the inverse problem is ill-posed
due to its potentially ambiguous solutions and its sensitivity to
the transport data. By training the neural network, Eq. \eqref{eq:neuralnet}
on cross sections from the LXCat project, we restrict the solution
to a subset of physically-plausible cross sections, thereby improving
the posedness of the inverse problem.}
\end{figure}

When the available swarm transport data is limited, the inverse swarm
problem becomes ill-posed with solutions that are non-unique and that
consist of cross sections that are sensitive to small variations in
the transport coefficients, as illustrated in Figure \ref{fig:Euler-diagram-illustrating}.
In such situations this lack of information can to some extent be
made up for by the experience and intuition of an expert. Due to this
required expertise, as well as the tedious nature of this trial and
error approach to swarm analysis, a number of automated methods for
solving the inverse swarm problem have been proposed \citep{Duncan1972,OMalley1980,Taniguchi1987,Morgan1991a,Morgan1991,Morgan1993,Brennan1993}.
Of particular relevance to the present work is an investigation in
the early 1990s by Morgan \citep{Morgan1991}, who solved the inverse
swarm problem using an artificial neural network that had been trained
on exemplar mappings between swarm data and cross sections. He concluded
that such a machine learning approach was well-suited for obtaining
an approximate cross section, that could subsequently be refined further
by nonlinear least-squares optimisation. Since Morgan’s pioneering
investigation, there have been major developments in the field of
machine learning that have allowed for much larger and more powerful
models to be realised \citep{Hinton2006,Hinton2006a,Bengio2007,Ranzato2007,Lee2008,Glorot2011}.
Given that there is now also a wealth of cross section data, made
available by the LXCat project \citep{Pancheshnyi2012,Pitchford2017,LXCat},
this data-driven approach to the inverse swarm problem warrants renewed
investigation.

The remainder of this paper is structured as follows. In the following
section, we present a new neural network for the solution of the inverse
swarm problem. In Section \ref{sec:Application-to-simulated}, we
apply this neural network to determine cross sections of helium and
argon using simulated swarm transport data. Finally, Section \ref{sec:Conclusion}
presents some conclusions and a discussion of future generalisations
and applications of this work.

\section{\label{sec:Neural-network-regression}Neural network regression of
cross sections}

Artificial neural networks are highly-parameterised mathematical functions
capable of universal function approximation \citep{Cybenko1989,Hornik1991}.
By carefully adjusting the parameters of a neural network, in a process
called \textit{training}, it is possible to approximate arbitrary
multivariate vector-valued functions. In other words, neural networks
are capable of performing nonlinear mappings between vector spaces.
In this section, we present a neural network for the solution of the
inverse swarm problem that learns from the inverse of the forward
mapping from cross sections to transport coefficients. This process
is schematically depicted in Figure \ref{fig:Euler-diagram-illustrating}.
Our neural network is of a similar architecture to that employed by
Morgan \citep{Morgan1991} and, as such, we will begin by outlining
the structure of Morgan's neural network before describing how ours
differs.

\subsection{\label{subsec:Model-overview}Model introduction and overview}

To perform the cross section regression given the transport coefficients,
we use a fully-connected neural network (FCNN). The simplest FCNN
is an affine transformation of an input vector $\mathbf{x}$ to an
output vector $\mathbf{y}$:
\begin{equation}
\mathbf{y}=\mathbf{W}\mathbf{x}+\mathbf{b},
\end{equation}
where the matrix $\mathbf{W}$ and vector $\mathbf{b}$ contain the
network parameters. More generally, to describe nonlinear relationships,
additional affine transformations are applied, each interleaved with
a nonlinear \textit{activation function}. For example, Morgan considered
an FCNN of the form \citep{Morgan1991}:
\begin{equation}
\mathbf{y}=\left(\mathbf{A}_{3}\circ\tanh\circ\mathbf{A}_{2}\circ\tanh\circ\mathbf{A}_{1}\right)\left(\mathbf{x}\right),
\end{equation}
where a hyperbolic tangent activation function is applied element-wise
throughout, and each affine transformation $\mathbf{A}_{n}\left(\mathbf{x}\right)=\mathbf{W}_{n}\mathbf{x}+\mathbf{b}_{n}$
has associated parameters $\mathbf{W}_{n}$ and $\mathbf{b}_{n}$.
This network is said to have $4$ \textit{layers} of \textit{neurons},
corresponding to the vectors of numbers reached at each stage of the
computation. For example, the above network has an input layer $\mathbf{x}$,
an output layer $\mathbf{y}$ and $2$ intermediate hidden layers
$\mathbf{x}_{1}$ and $\mathbf{x}_{2}$, which arise as follows:
\begin{eqnarray}
\mathbf{x}_{1} & = & \tanh\left(\mathbf{A}_{1}\left(\mathbf{x}\right)\right),\\
\mathbf{x}_{2} & = & \tanh\left(\mathbf{A}_{2}\left(\mathbf{x}_{1}\right)\right),\\
\mathbf{y} & = & \mathbf{A}_{3}\left(\mathbf{x}_{2}\right).
\end{eqnarray}
While the input and output layers, $\mathbf{x}$ and $\mathbf{y}$,
are generally of a size fixed by the nature of the problem at hand,
the number and size of the hidden layers must be specified. With too
few hidden neurons, the predictive power of the model is hindered.
Conversely, a model that is too complex can overspecialise to the
training data, resulting in poor generalisation to gases outside of
the training dataset. Fortunately, this latter problem becomes less
likely to occur as the amount of training data increases.

The main architectural difference between our network and Morgan's
is the structure of the input and output vectors, $\mathbf{x}$ and
$\mathbf{y}$. Morgan considered an input vector $\mathbf{x}$ containing
measured drift velocities $W$ and characteristic energies $D_{T}/\mu$,
where $D_{T}$ is the bulk transverse diffusion coefficient and $\mu$
is the bulk electron mobility, and an output vector $\mathbf{y}$
as a discrete representation of the cross section of interest, $\sigma\left(\varepsilon\right)$
\citep{Morgan1991}:
\begin{equation}
\mathbf{x}=\left[\begin{array}{c}
W\left(E_{1}/n_{0}\right)\\
D_{T}\left(E_{1}/n_{0}\right)/\mu\left(E_{1}/n_{0}\right)\\
W\left(E_{2}/n_{0}\right)\\
D_{T}\left(E_{2}/n_{0}\right)/\mu\left(E_{2}/n_{0}\right)\\
\vdots
\end{array}\right],\quad\mathbf{y}=\left[\begin{array}{c}
\sigma\left(\varepsilon_{1}\right)\\
\sigma\left(\varepsilon_{2}\right)\\
\vdots
\end{array}\right].\label{eq:morganvectors}
\end{equation}
Here, the input transport coefficient measurements are performed at
various reduced electric fields $E_{1}/n_{0},E_{2}/n_{0},\dots$,
where $n_{0}$ is the number density of the background neutrals, and
the output cross section is sampled at discrete energies $\varepsilon_{1},\varepsilon_{2},\dots$.
As input in our case, in addition to the transport coefficients of
bulk drift velocity $W$ and bulk longitudinal diffusion $n_{0}D_{L}$,
we also consider the first Townsend ionisation coefficient $\alpha/n_{0}$.
For the output cross section, rather than discretising as above, we
use the neural network itself as a discrete approximation of the cross
section as a function of energy. Energy, $\varepsilon$, now becomes
an input to the neural network alongside the transport coefficients,
and the output is now a single number corresponding to the cross section
evaluated at this energy, $\sigma\left(\varepsilon\right)$:
\begin{equation}
\mathbf{x}=\left[\begin{array}{c}
\varepsilon\\
W\left(E_{1}/n_{0}\right)\\
n_{0}D_{L}\left(E_{1}/n_{0}\right)\\
\alpha\left(E_{1}/n_{0}\right)/n_{0}\\
W\left(E_{2}/n_{0}\right)\\
n_{0}D_{L}\left(E_{2}/n_{0}\right)\\
\alpha\left(E_{2}/n_{0}\right)/n_{0}\\
\vdots
\end{array}\right],\quad y=\sigma\left(\varepsilon\right).
\end{equation}
Although it is not shown here, before being used for regression all
the data is log-transformed with Eq. \eqref{eq:log-transform}. This
is detailed in the following section.

\begin{figure}
\begin{centering}
\includegraphics[scale=0.55]{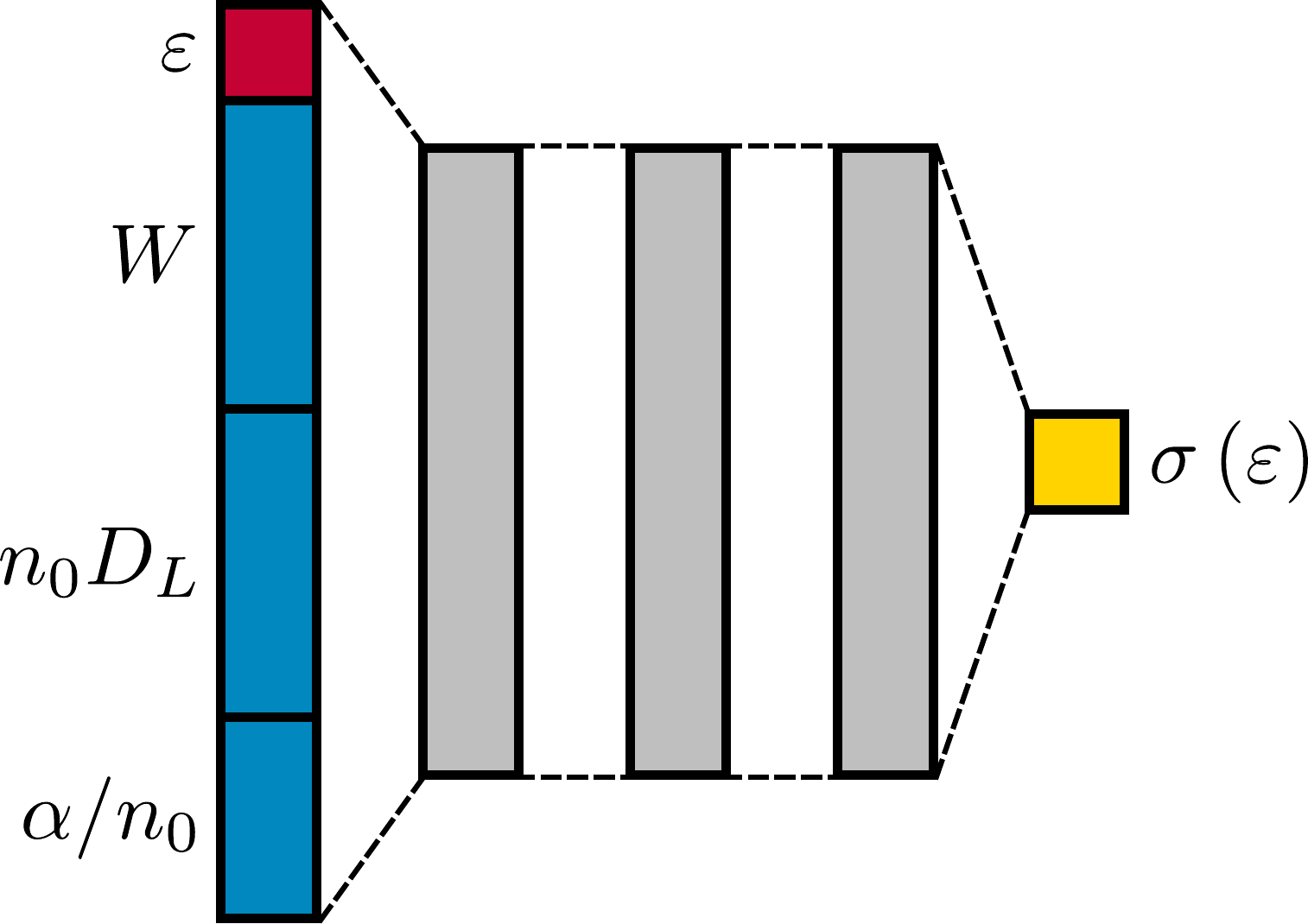}
\par\end{centering}
\caption{\label{fig:Diagram-(not-to}Diagram (not to scale) of the fully-connected
neural network used for regression of a cross section (yellow) as
a function of energy (red), given an associated set of transport coefficients
(blue). The number and size of intermediate hidden layers (grey) affects
the capacity of the model to perform this nonlinear mapping. For the
cross section fits in Section \ref{sec:Application-to-simulated},
$3$ hidden layers were used, each containing $100$ neurons. For
the simultaneous fitting of multiple cross sections in Section \ref{subsec:Fitting-multiple-cross},
additional output neurons were added accordingly.}
\end{figure}

This architectural change of making the energy an explicit input,
illustrated in Figure \ref{fig:Diagram-(not-to}, provides a number
of benefits. Firstly, rather than having to decide on a one-size-fits-all
cross section discretisation, the neural network learns a suitable
discretisation that is tailored to the training data. Secondly, now
that there is only one output neuron per cross section, far fewer
weights are required in the final affine transformation, making the
network simpler to train, faster to evaluate, and less likely to overspecialise
to the training data. This is most beneficial when multiple output
cross sections are considered, such as in Section \ref{subsec:Fitting-multiple-cross}.
Finally, it is now relatively trivial to constrain the cross section
training data when there are known upper and lower bounds obtained
from, for example, crossed beam experiments \citep{Filippelli1994}.
This is awkward to do when simultaneously predicting a cross section
at multiple energies, as in Eq. \eqref{eq:morganvectors}, because
then constraints must also be enforced simultaneously for every energy
considered.

Our neural network, as used in Section \ref{sec:Application-to-simulated}
takes the form:
\begin{equation}
\sigma\left(\mathbf{x}\right)=\left(A_{4}\circ\mathrm{swish}\circ\mathbf{A}_{3}\circ\mathrm{swish}\circ\mathbf{A}_{2}\circ\mathrm{swish}\circ\mathbf{A}_{1}\right)\left(\mathbf{x}\right),\label{eq:neuralnet}
\end{equation}
where $3$ hidden layers of $100$ neurons each have been used, and
the ``swish'' activation function has been applied element-wise
throughout, where $\mathrm{swish}\left(x\right)=x/\left(1+e^{-x}\right)$
\citep{Ramachandran2017}. It is likely that there are different,
more optimal, choices for these neural network \textit{hyperparameters},
however determining these is beyond the scope of this investigation.
Instead, we make sure to train the network using a large amount of
data, so as to discourage it from overspecialising in the event that
we have specified a network that is larger than is ideal.

\subsection{\label{subsec:Data-preparation-and}Data preparation and model training}

To train the neural network, Eq. \eqref{eq:neuralnet}, we require
example inputs and outputs, a cost function that indicates the performance
of the neural network on this training data, and an optimisation algorithm
to minimise this cost function by varying the parameters of the neural
network.

\begin{figure}
\begin{centering}
\includegraphics[scale=0.65]{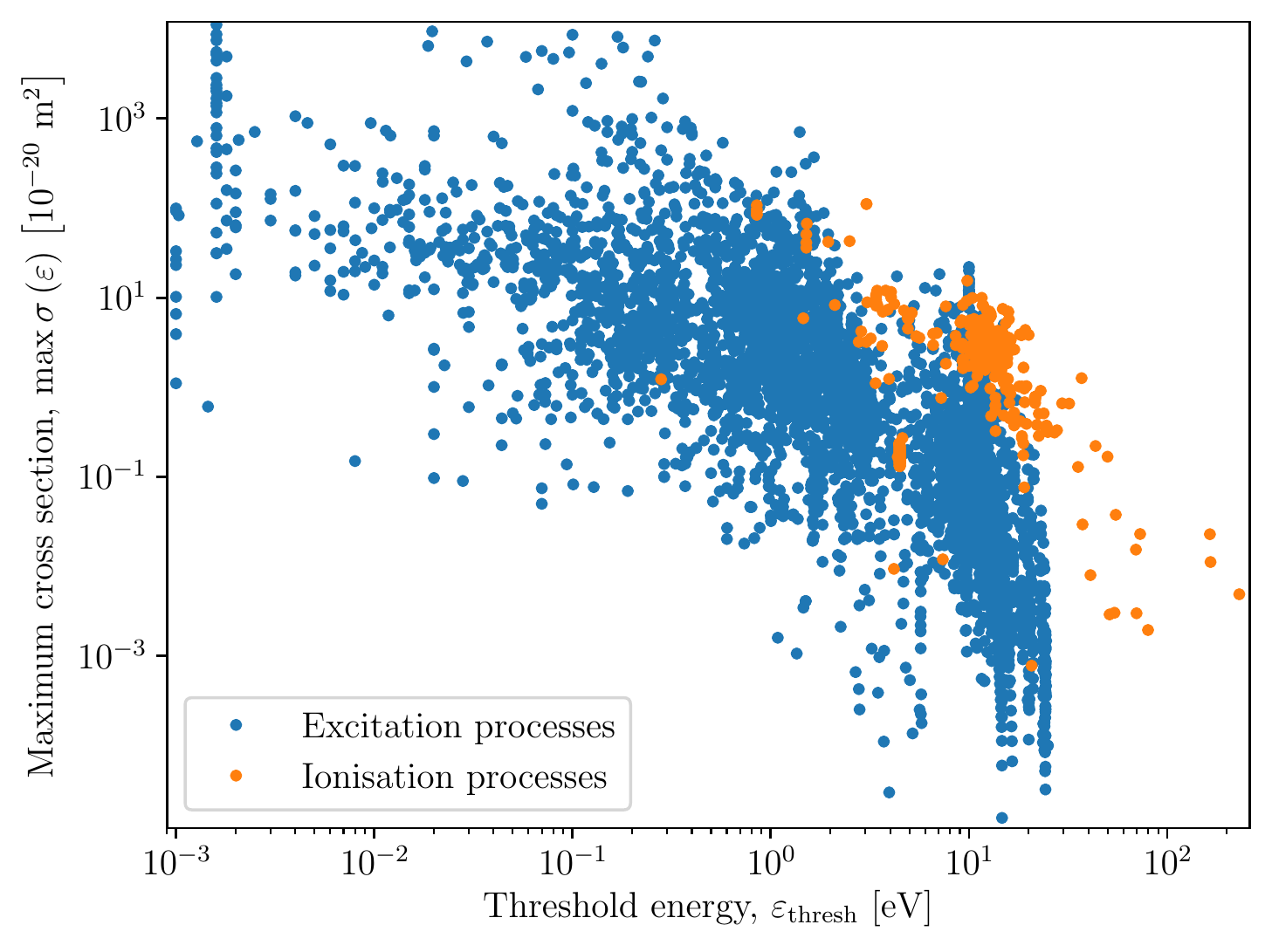}
\par\end{centering}
\caption{\label{fig:A-scatter-plot}Scatter plot illustrating the rough power-law
relationship between the magnitude and threshold energy of the excitation
and ionisation cross sections on LXCat. Pearson's correlation coefficient
between the log-transformed maximum cross section and log-transformed
threshold energy is $r=-0.738$ for excitation processes and $r=-0.755$
for ionisation processes. By training the neural network, Eq. \eqref{eq:neuralnet},
on cross sections sampled using Eq. \eqref{eq:mixture}, we ensure
this negative correlation is reflected in the training data.}
\end{figure}

We consider first the issue of determining suitable cross sections
for the training. For the purpose of fitting elastic momentum transfer
cross sections, Morgan generated training examples using a power-law
model of the form $\sigma\left(\varepsilon\right)\propto\varepsilon^{p}$,
where $-1\leq p\leq1$ \citep{Morgan1991}. Similar options exist
for other processes, e.g. Machacek \textit{et al}. described the cross
section for total positronium formation by positron scattering using
a surge function $\sigma\left(\varepsilon\right)\propto\varepsilon^{p}e^{-b\varepsilon}$,
where $p$ and $b$ are shape parameters \citep{Machacek2016}. Parameterised
cross sections such as these are ideal for machine learning, as they
can be sampled from indefinitely for training data. That said, we
choose instead to train using a finite set of cross sections from
the LXCat project \citep{Pancheshnyi2012,Pitchford2017,LXCat}, so
as to expose the neural network to as much measured and calculated
cross section physics as possible. To make up for this finite amount
of cross section data, we generate training data by random interpolation
of LXCat cross sections. Specifically, given a unique pair of cross
sections, $\sigma_{1}\left(\varepsilon\right)$ and $\sigma_{2}\left(\varepsilon\right)$,
as well as a uniformly distributed mixing ratio $r\in\left[0,1\right]$,
we form each training cross section by evaluating:
\begin{equation}
\sigma\left(\varepsilon\right)=\sigma_{1}^{1-r}\left(\varepsilon+\varepsilon_{1}-\varepsilon_{1}^{1-r}\varepsilon_{2}^{r}\right)\sigma_{2}^{r}\left(\varepsilon+\varepsilon_{2}-\varepsilon_{1}^{1-r}\varepsilon_{2}^{r}\right),\label{eq:mixture}
\end{equation}
where $\varepsilon_{1}$ and $\varepsilon_{2}$ are the respective
threshold energies of $\sigma_{1}\left(\varepsilon\right)$ and $\sigma_{2}\left(\varepsilon\right)$.
This is a weighted geometric average of the two cross sections that
has been shifted to have a new threshold of $\varepsilon_{1}^{1-r}\varepsilon_{2}^{r}$,
which is itself a weighted geometric average of the separate threshold
energies. We use the same ratio $r$ when mixing both cross sections
and threshold energies, as there is a negative correlation between
the magnitude of a cross section and its threshold energy, depicted
in Figure \ref{fig:A-scatter-plot}, that we would like to see reflected
in the training data.

Each cross section generated using Eq. \eqref{eq:mixture} must be
sampled at discrete energies in order to be used for training. To
aid the neural network in learning the global energy-dependence of
the cross sections, we sample at a large number of energies over the
domain of interest. These energies are also selected geometrically:
\begin{equation}
\varepsilon=\varepsilon_{\mathrm{min}}^{1-s}\varepsilon_{\mathrm{max}}^{s},\label{eq:sample}
\end{equation}
where $\left[\varepsilon_{\mathrm{min}},\varepsilon_{\mathrm{max}}\right]$
is the energy domain under consideration and $s\in\left[0,1\right]$
is a uniformly distributed random number.

Finally, from our chosen cross section or cross sections, we must
determine corresponding transport coefficients to complete our input/output
training pair. For efficient and robust generation of this swarm data,
here we apply the two-term approximation \citep{Hagelaar2005,Robson1986}
to Boltzmann's equation \eqref{eq:BE} and then perform backward prolongation
\citep{Sherman1960} of the EEDF by inward integration from high to
low energies. This integration is performed numerically, using an
adaptive order adaptive energy Adams-Moulton method \citep{Hairer1993},
as implemented in the \textit{DifferentialEquations.jl} software ecosystem
\citep{Rackauckas2017,DelayDiffEq,OrdinaryDiffEq}.

Since cross sections, energies and transport coefficients all span
many orders of magnitude, we make sure to log-transform everything
before training. That is, given a strictly positive quantity $z$,
we apply a log-transformation that is scaled and shifted to ensure
all the training data lies within the domain $\left[-1,1\right]$:
\begin{equation}
z\mapsto\log_{\sqrt{\frac{z_{\mathrm{max}}}{z_{\mathrm{min}}}}}\left(\frac{z}{\sqrt{z_{\mathrm{max}}z_{\mathrm{min}}}}\right),\label{eq:log-transform}
\end{equation}
where $z_{\min}$ and $z_{\mathrm{max}}$ are the extrema of all instances
of $z$ in the training dataset. If any instance of $z$ happens to
be zero, we replace it with a suitably small positive number before
applying the above transformation.

To train the neural network, we minimise its mean absolute error on
the training set of cross sections:
\begin{equation}
\frac{1}{N}\sum_{i=1}^{N}\left|y_{i}-\sigma\left(\mathbf{x}_{i}\right)\right|,\label{eq:MAE}
\end{equation}
where the index $i$ ranges over the entire set of $N$ training examples
$\left(\mathbf{x}_{i},y_{i}\right)$, and $\sigma\left(\mathbf{x}_{i}\right)$
is the corresponding neural network prediction for each. We have purposefully
avoided using the mean squared error, as that encourages the neural
network to preferentially improve upon its worst predictions. This
sounds reasonable, and is the motivation behind least-squares regression,
but as we are solving an inverse problem there is an inherent uncertainty
to the solution that fundamentally limits how well the network can
perform. Encouraging the neural network to do the impossible of accurately
and consistently fitting the least certain cross section values is
expected to hinder its performance overall.

\begin{figure}
\begin{centering}
\includegraphics[scale=0.65]{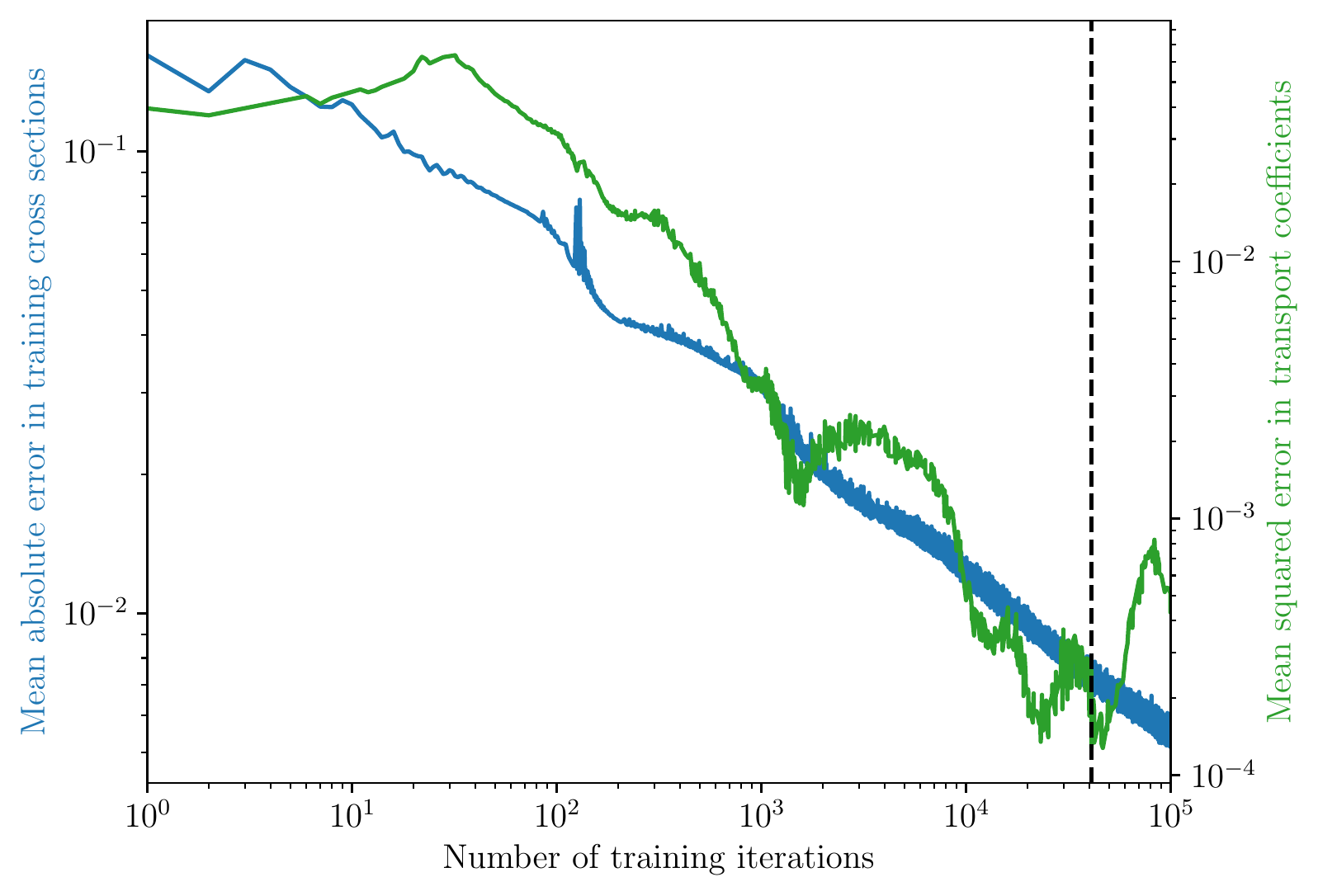}
\par\end{centering}
\caption{\label{fig:Performance-of-the}Mean absolute error (blue), Eq. \eqref{eq:MAE},
of the neural network on the training set used to fit argon's elastic
momentum transfer cross section in Figure \ref{fig:Neural-network-fit-argon-elastic},
alongside the mean squared error (green) in reproducing the transport
coefficients used to perform the fit. As training progresses, the
error for the training dataset decreases as expected. Beyond a certain
point, however, the neural network overspecialises to the training
data, at the expense of the argon fit. We conclude that the ideal
place to stop training is where the mean squared error in the transport
coefficients is minimised, indicated here by the vertical dashed line.}
\end{figure}

We implement and train the model using the \textit{Flux.jl} machine
learning framework \citep{Innes2018}. In constructing the neural
network, we initialise the parameters in $\mathbf{b}_{n}$ to zero
and those in $\mathbf{W}_{n}$ to uniform random numbers as described
by Glorot and Bengio \citep{Glorot2010}. Then, we train by minimising
Eq. \eqref{eq:MAE} using the Adam optimiser \citep{Kingma2015} with
step size $\alpha=10^{-3}$ and exponential decay rates $\beta_{1}=0.9$
and $\beta_{2}=0.999$. By definition, training improves the performance
of the neural network on the training examples, with the hope that
good performance on the training set will be correlated with good
performance in general. However, with too much training there lies
a danger of the neural network learning features that are unique to
the training data, at the expense of generalisation \citep{Geman1992}.
A common technique to avoid this is to simply stop the training before
the network's ability to generalise worsens, as quantified by some
other measure independent of the data used for training \citep{Morgan1990}.
For this measure, we use the necessary condition that the solution
to the inverse swarm problem must also satisfy the forward swarm problem.
Thus, we continually apply the Boltzmann solver, while training, to
determine how well the fitted cross sections reproduce the transport
coefficients that were used to perform the fit. We then stop training
when the mean squared error in this reproduced swarm data reaches
a minimum, as demonstrated in Figure \ref{fig:Performance-of-the}.

\section{\label{sec:Application-to-simulated}Application to simulated swarm
data}

In this section, we simulate pulsed Townsend swarm experiments for
electron transport in helium and argon at a temperature $T=300\ \mathrm{K}$,
across a range of $50$ reduced electric fields $E/n_{0}$ spaced
exponentially between $10^{-3}\ \mathrm{Td}$ and $10^{3}\ \mathrm{Td}$
inclusive, where $1\ \mathrm{Td}=1\ \mathrm{Townsend}=10^{-21}\ \mathrm{V\ m^{2}}$.
Using the resulting transport coefficients of bulk drift velocity
$W$, bulk longitudinal diffusion coefficient $n_{0}D_{L}$ and Townsend
coefficient $\alpha/n_{0}$ as inputs, we then apply the neural network
Eq. \eqref{eq:neuralnet} toward fitting various cross sections of
helium and argon over the energy domain $\left[10^{-1}\ \mathrm{eV},10^{2}\ \mathrm{eV}\right]$.
For both gases, cross section data are sourced from the Biagi v7.1
database \citep{Biagiv71}.

\subsection{\label{subsec:Fitting-elastic-cross}Fitting elastic momentum transfer
cross sections}

\begin{figure}
\begin{centering}
\includegraphics[scale=0.65]{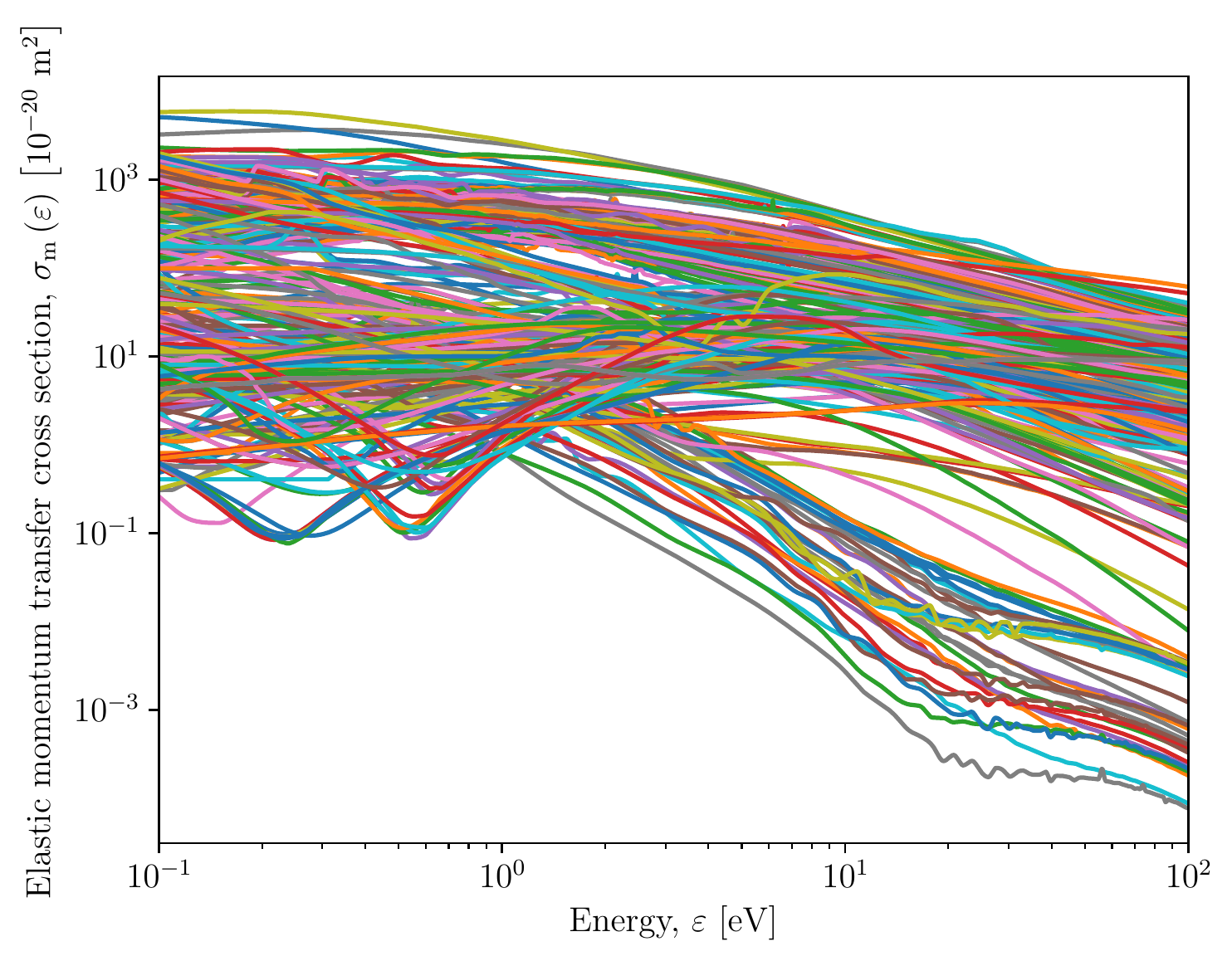}
\par\end{centering}
\caption{\label{fig:All-elastic-cross}All currently available elastic momentum
transfer cross sections from the LXCat project \citep{Biagi,Biagiv71,Bordage,BSR,CCC,Christophorou,COP,FLINDERS,Hayashi,ISTLisbon,Itikawa,Morgan,Puech,QUANTEMOL,SIGLO,TRINITI,BiagiMAGBOLTZ,Zatsarinny2004,Allan2006,Bray1992,Fursa1995,Zammit2014,Zammit2016,Christophorou2000,Christophorou2000a,Alves2014,Quantemolwebsite}.
These are randomly sampled from using Eq. \eqref{eq:mixture} and
the resulting cross sections are employed in the training of the neural
network depicted in Figure \ref{fig:Diagram-(not-to} for the purpose
of fitting elastic momentum transfer cross sections. A similar approach
is taken when fitting cross sections for other collision processes.}
\end{figure}

\begin{figure}
\begin{centering}
\includegraphics[scale=0.52]{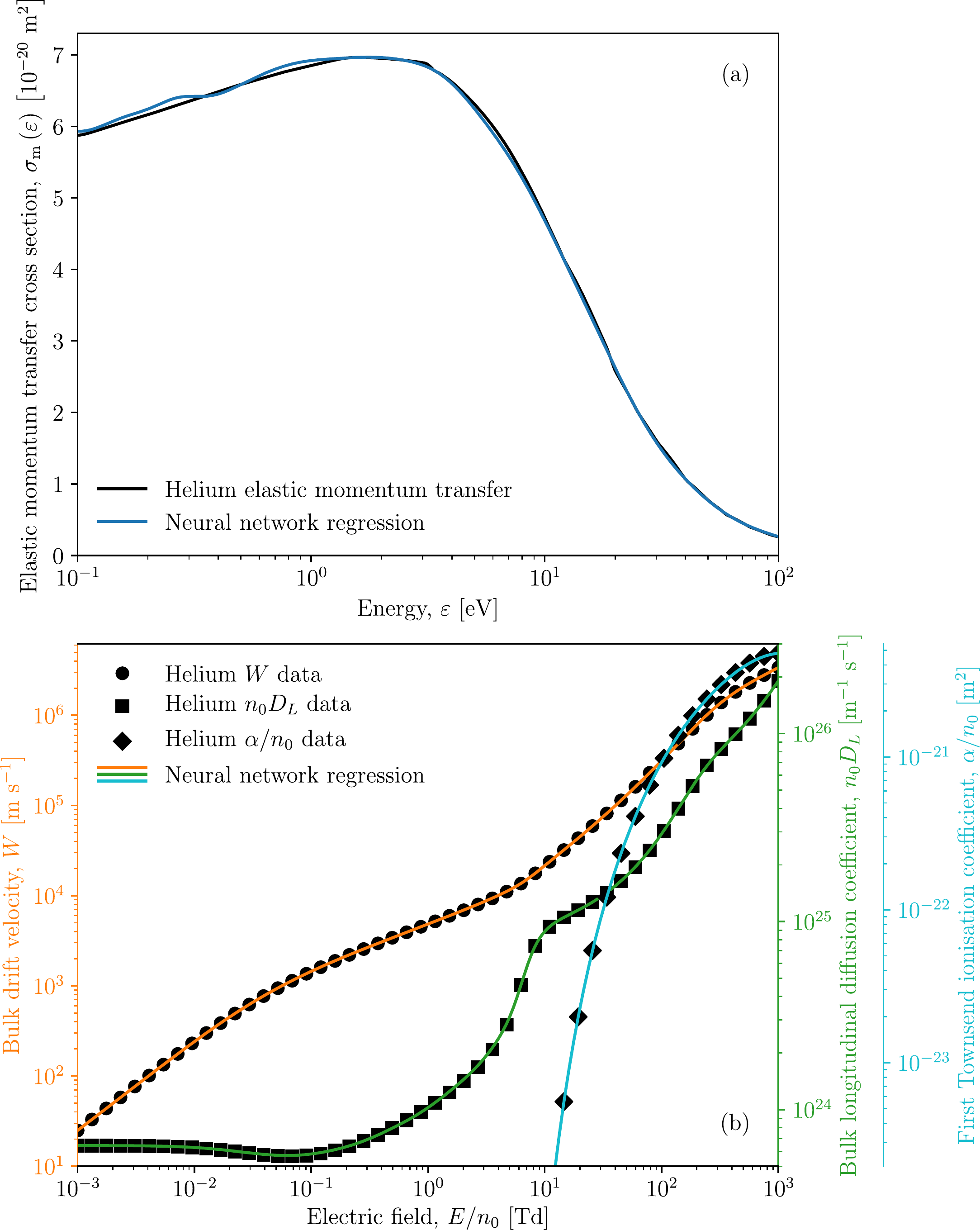}
\par\end{centering}
\caption{\label{fig:Neural-network-fit-He-elastic}Neural network regression
of helium's elastic MTCS, (a), alongside corresponding plots of the
transport coefficients, (b). The neural network determines the cross
section to within an accuracy of $4\%$ here, while also being consistent
with the provided swarm data to within the same margin of error.}
\end{figure}

We consider first fitting helium's elastic momentum transfer cross
section (MTCS), while assuming its inelastic cross sections are known.
For the training data, we use Eq. \eqref{eq:mixture} to randomly
sample $10^{4}$ elastic MTCS from those plotted in Figure \ref{fig:All-elastic-cross}
from LXCat \citep{Biagi,Biagiv71,Bordage,BSR,CCC,Christophorou,COP,FLINDERS,Hayashi,ISTLisbon,Itikawa,Morgan,Puech,QUANTEMOL,SIGLO,TRINITI,BiagiMAGBOLTZ,Zatsarinny2004,Allan2006,Bray1992,Fursa1995,Zammit2014,Zammit2016,Christophorou2000,Christophorou2000a,Alves2014,Quantemolwebsite}.
Of course, helium is excluded when performing this sampling. For each
of these generated cross sections, corresponding swarm transport coefficients
are found. Finally, each cross section is sampled at $100$ random
energies between $10^{-1}\ \mathrm{eV}$ and $10^{2}\ \mathrm{eV}$,
using Eq. \eqref{eq:sample}, resulting in a total of $10^{6}$ training
examples. We split these training exemplars into $10$ batches of
$10^{5}$ examples each, making sure to shuffle the training data
beforehand so that each batch is representative of the training set
as a whole. Training then proceeds iteratively by repeatedly cycling
through each batch in turn, updating the neural network parameters
each time using the optimiser. Training is continued until the mean
squared error in the resulting transport coefficients reaches a minimum,
as demonstrated in Figure \ref{fig:Performance-of-the}. The final
result of this process, for the fitted cross section and corresponding
transport  coefficients, is plotted in Figure \ref{fig:Neural-network-fit-He-elastic}.
The neural network is found to be accurate here to within $4\%$,
for both the elastic MTCS and its corresponding transport coefficients.

\begin{figure}
\begin{centering}
\includegraphics[scale=0.52]{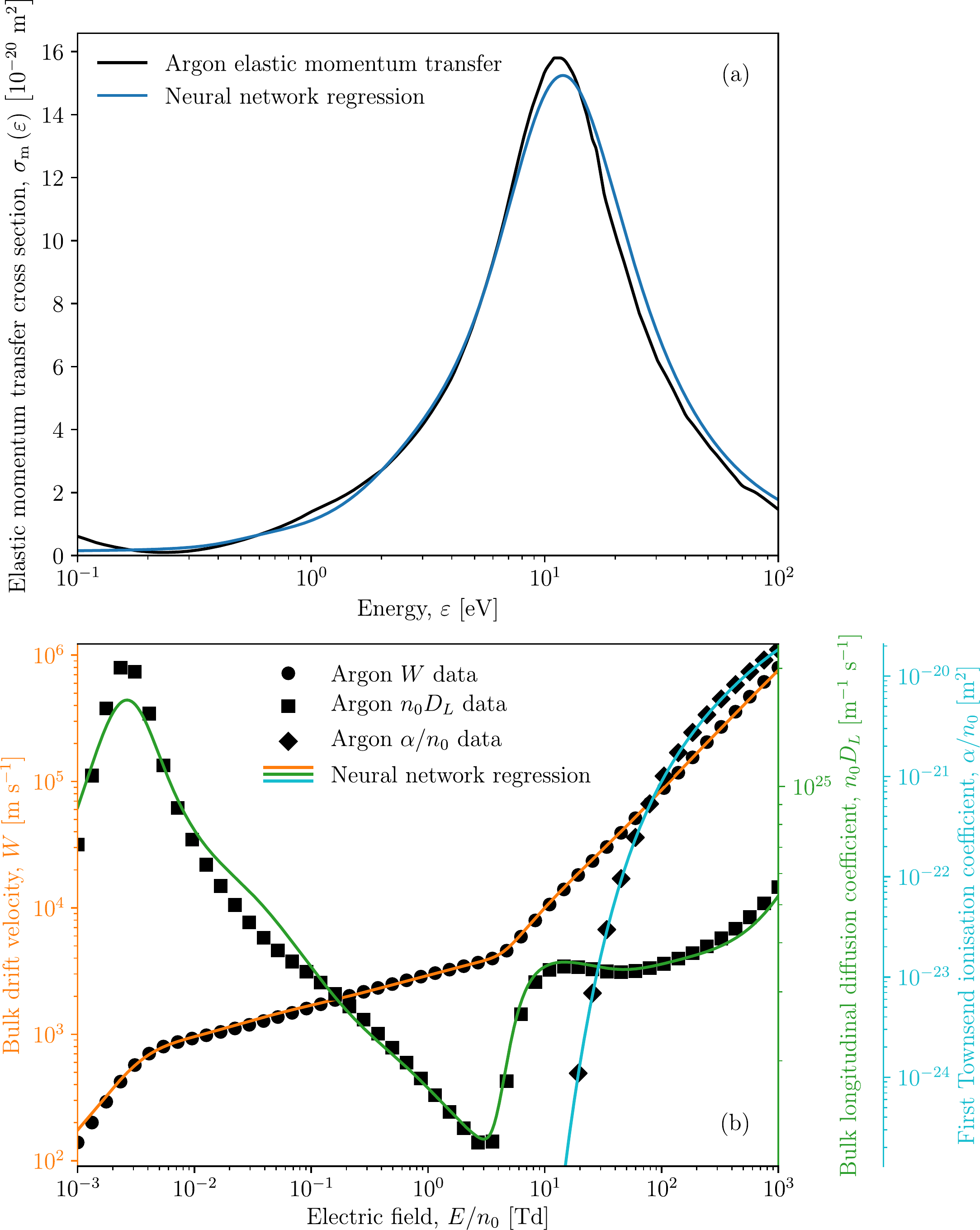}
\par\end{centering}
\caption{\label{fig:Neural-network-fit-argon-elastic}Neural network regression
of argon's elastic MTCS, (a), alongside corresponding plots of the
transport coefficients, (b). Due to the uniqueness of argon's Ramsauer-Townsend
minimum, the training data, plotted in Figure \ref{fig:All-elastic-cross},
is insufficient to perform an accurate cross section fit. Away from
that minimum, the fit is accurate to within $20\%$. It is expected
that the fit quality would improve if more suitable cross sections
were to be available and thus used for training.}
\end{figure}

We now repeat the above process for argon (whose elastic MTCS is now
excluded from the training data instead of helium's) and plot the
results in Figure \ref{fig:Neural-network-fit-argon-elastic}. Unlike
helium, argon's elastic MTCS is an outlier among the training data,
with its Ramsauer-Townsend minimum dipping below the rest of the LXCat
cross sections in Figure \ref{fig:All-elastic-cross}. Because of
this, the neural network struggles to determine the correct magnitude
of that minimum, believing it to be more than twice as large as it
should. In effect, our choice of training data has incorrectly constrained
the fitted cross section. The error at larger energies, above $0.4\ \mathrm{eV}$,
is however fortunately considerably better, lying to within $20\%$.
This larger error, as compared to helium, is likely due in part to
the neural network compensating for the incorrect Ramsauer-Townsend
minimum, so as to better match the transport coefficients. Lastly,
it should be noted that a larger error is to be expected for argon.
This follows as the chosen swarm data provides less information about
argon's elastic MTCS than it does for helium's, as evidenced by the
mean energy of electrons at the highest applied field, $E/n_{0}=1000\ \mathrm{Td}$,
which is $\sim133\ \mathrm{eV}$ for helium and only $\sim15\ \mathrm{eV}$
for argon.

\subsection{\label{subsec:Fitting-threshold-cross}Fitting threshold cross sections}

\begin{figure}
\begin{centering}
\includegraphics[scale=0.52]{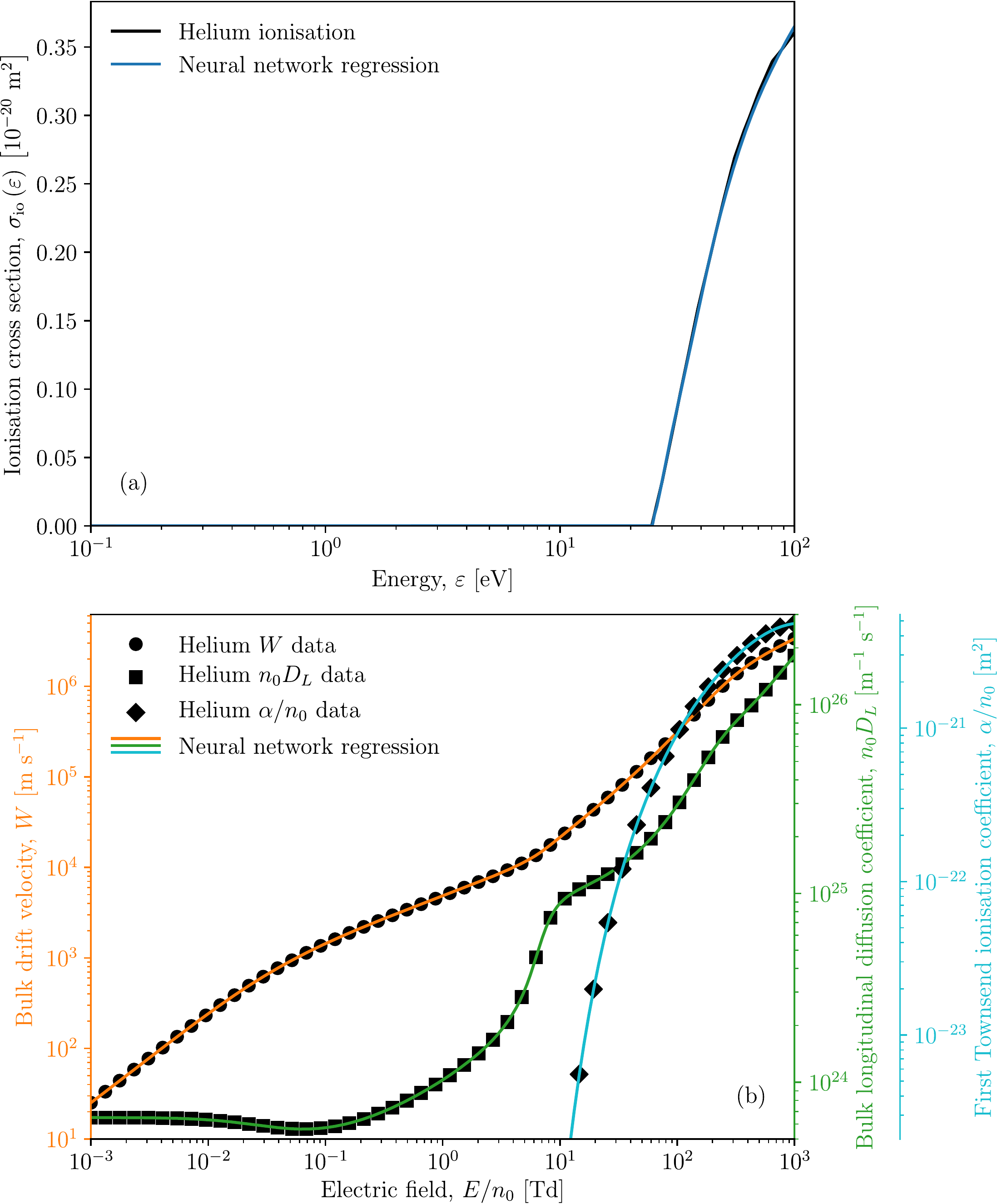}
\par\end{centering}
\caption{\label{fig:Neural-network-fit-helium-ionisation}Neural network regression
of helium's ionisation cross section, (a), alongside corresponding
plots of the transport coefficients, (b). The accuracy here is comparable
to that seen for the elastic MTCS fit depicted in Figure \ref{fig:Neural-network-fit-He-elastic}.}
\end{figure}

We turn now to fitting the ionisation cross section of helium, as
a representative example of data with a clear threshold energy. We
generate our training data using LXCat as before, only now employing
the ionisation cross sections instead \citep{Biagi,Biagiv71,Bordage,BSR,CCC,Christophorou,Hayashi,ISTLisbon,Itikawa,Morgan,Phelps,Puech,QUANTEMOL,SIGLO,TRINITI,BiagiMAGBOLTZ,Zatsarinny2004,Allan2006,Bray1992,Fursa1995,Zammit2014,Zammit2016,Christophorou2000,Christophorou2000a,Alves2014,Phelpswebsite,Quantemolwebsite}.
Fortunately, besides from this change in the utilised cross section
data, no other alterations are required to the neural network or training
procedure in order to fit a threshold cross section. Note that we
determine the corresponding threshold energy by using a cross section
threshold of $10^{-25}\ \mathrm{m}^{2}$, below which the cross section
is taken as being equal to zero. Figure \ref{fig:Neural-network-fit-helium-ionisation}
plots the result of the neural network regression, alongside corresponding
transport coefficients. Above threshold, the fit of helium's ionisation
cross section is in error by at most $4\%$, hence on a par with the
previous fit of helium's elastic MTCS. It is very promising to see
that the neural network is able to successfully identify the ionisation
threshold from the transport data. Machine learning approaches thus
have the potential for determining unknown threshold energies, such
as the neutral dissociation threshold of tetrahydrofurfuryl alcohol
(THFA), a molecule of biological interest \citep{Jones2013,Limao-Vieira2014,Duque2014a,Mozejko2006,Bellm2012,Duque2014,Chiari2014,Brunger2017}.

\subsection{\label{subsec:Fitting-multiple-cross}Fitting multiple cross sections
simultaneously}

Finally, we consider simultaneously fitting helium's elastic momentum
transfer, $n=2$ singlet excitation ($2^{1}S$ and $2^{1}P$), $n=2$
triplet excitation ($2^{3}S$ and $2^{3}P$) and ionisation cross
sections. Accordingly, we now also incorporate excitation cross section
data from LXCat into the training procedure \citep{Biagi,Biagiv71,Bordage,BSR,CCC,Christophorou,FLINDERS,Hayashi,ISTLisbon,Itikawa,Morgan,NGFSRDW,Phelps,Puech,SIGLO,TRINITI,BiagiMAGBOLTZ,Zatsarinny2004,Allan2006,Bray1992,Fursa1995,Zammit2014,Zammit2016,Christophorou2000,Alves2014,Phelpswebsite}.
As in the previous fits, we assume nothing here but the number and
type of each cross section. Correspondingly, the output layer of our
neural network is now increased to contain $4$ neurons, one for each
cross section being fitted. How these cross sections are ordered in
the output layer does not matter, so long as the ordering is kept
consistent when training and applying the network. With this in mind,
note that it is not clear how to consistently order each training
pair of $n=2$ excitation cross sections. This is because excitation
processes can be interchanged with one another without any affect
on the transport coefficients. If we are not careful, the neural network
could proceed to correctly predict the excitation cross sections of
helium, but be penalised for doing so because the cost function, Eq.
\eqref{eq:MAE}, happens to assume the opposite ordering. As such,
for this fit, we modify the cost function so as to make it symmetric
with respect to the excitation cross sections. In effect, for each
training exemplar, we try both permutations of predicted excitation
cross sections and select the one that minimises the cost function
the most.

\begin{figure}
\begin{centering}
\includegraphics[scale=0.52]{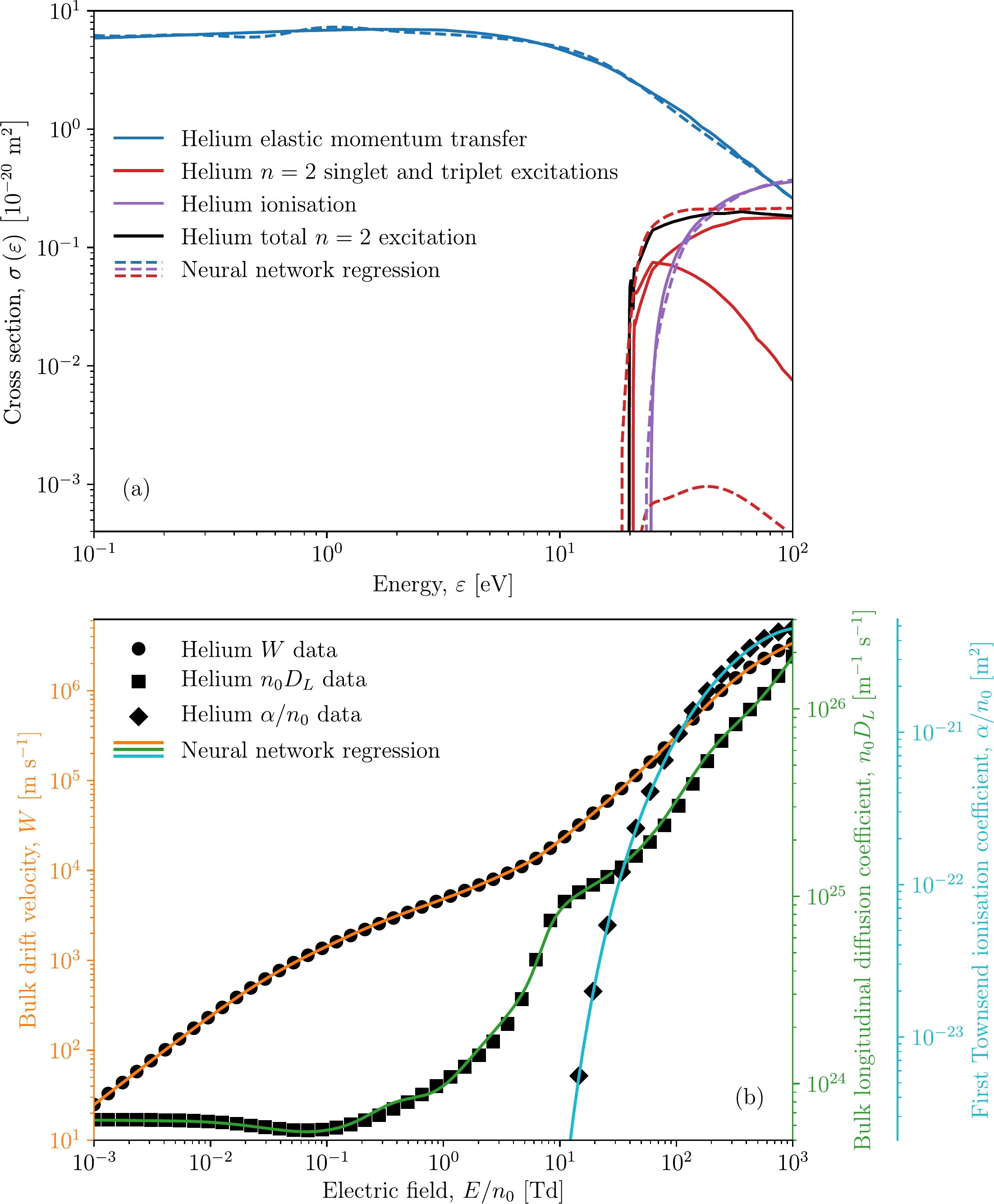}
\par\end{centering}
\caption{\label{fig:Neural-network-fit-helium-lots}Neural network regression
of helium's elastic momentum transfer, $n=2$ singlet excitation,
$n=2$ triplet excitation and ionisation cross sections, (a), alongside
corresponding plots of the transport coefficients, (b). As expected,
the overall error has increased here compared to fitting helium's
elastic momentum transfer and ionisation cross sections individually.
The network was unable to unfold the separate cross sections for $n=2$
singlet and triplet excitation, although the values it predicts are
consistent with the threshold and magnitude of the total $n=2$ excitation
cross section.}
\end{figure}

Figure \ref{fig:Neural-network-fit-helium-lots} plots the result
of the neural network regression for this case. Here we observe the
fitted elastic MTCS to be accurate to within roughly $10\%$ over
the entire range of energies. This is an expected increase over the
maximum $4\%$ error observed in Section \ref{subsec:Fitting-elastic-cross},
where the elastic MTCS was fitted exclusively. For inelastic processes
above threshold, we see larger cross section errors of within $20\%$
for total $n=2$ excitation and $25\%$ for ionisation. Of the pair
of $n=2$ excitation cross sections determined by the network, one
was comparable to the total $n=2$ excitation cross section, while
the other was made very small. This inability of the network to unfold
the separate $n=2$ singlet and $n=2$ triplet excitation cross sections
is likely due in part to their very similar threshold energies. Overall,
it is expected that the ability of the neural network to accurately
unfold multiple cross sections would improve with the consideration
of additional swarm data, such as transport coefficients of helium
mixed with other gases.

\section{\label{sec:Conclusion}Conclusion}

We have presented a machine learning approach to the inverse swarm
problem where a neural network is used to approximate unknown cross
sections as a functions of energy, given corresponding swarm transport
data. By training this network on physical cross sections from the
LXCat project \citep{Pancheshnyi2012,Pitchford2017,LXCat}, it was
found to yield physically-plausible solutions to the inverse swarm
problem that are consistent with the specified transport coefficients.
As a demonstration, we applied this network to determine cross sections
of helium and argon from simulated pulsed Townsend electron swarm
data over a range of reduced electric fields $E/n_{0}$. We found
that a suitably-trained neural network could determine individual
elastic momentum transfer and ionisation cross sections to within
$4\%$ accuracy for the swarm data considered. From the same swarm
data, we were also able to simultaneously fit the elastic MTCS, total
excitation, and ionisation cross sections of helium to within $10\%$,
$20\%$ and $25\%$ accuracy respectively. We were not able to resolve
the structure of the individual $n=2$ singlet and triplet excitation
cross sections of helium, as was in line with our expectations given
the similar threshold energies of these processes. Promisingly, this
data-driven approach to swarm analysis not only avoids the tedium
of conventional iteration, but also appears to display some of the
intuition of a domain expert.

A fundamental limitation of our neural network is that it provides
only a single solution to a problem for which many plausible solutions
may exist. Ideally, we would like to be able to quantify this uncertainty
in the solution and place corresponding error bars on the predicted
cross sections. One such approach involves determining a probability
distribution of cross sections, from which plausible solutions can
be sampled. This process is known as conditional density estimation
and a number of machine learning models are capable of performing
it, including mixture density networks \citep{Bishop1994}, conditional
variational autoencoders \citep{Sohn2015}, conditional generative
adversarial networks \citep{Mirza2014} and conditional flow-based
generative models \citep{Dinh2014,Dinh2016,Kingma2018}. Work on addressing
this present limitation in our approach is currently underway.

Finally, before our neural network can be applied to actual swarm
measurements, the presence of experimental error in these measurements
must be accounted for. Fortunately, by simply introducing simulated
experimental error estimates in the transport coefficients used for
the training, we should expect the neural network to adapt accordingly.
Indeed, this is the principle behind neural network approaches to
noise reduction in images \citep{Goodfellow2016}. This issue will
be addressed in our future studies.

\ack{}{}

The authors gratefully acknowledge the financial support of the Australian
Research Council through the Discovery Projects Scheme (Grant \#DP180101655).

\appendix

\bibliographystyle{iopart-num}
\addcontentsline{toc}{section}{\refname}\bibliography{references}

\end{document}